
%
%
%
%
%
%
%
%
%
%
%
%
\input harvmac
\def\Titletwo#1#2#3{\nopagenumbers\abstractfont\hsize=\hstitle\rightline{#1}%
\vskip 1in\centerline{\titlefont #2}\vskip .1in\centerline{\titlefont #3}
\abstractfont\vskip .5in\pageno=0}
\Titletwo{FERMI-PUB-92/169-T}{String Theory, Black Holes, and}{SL(2,R) Current
Algebra}

\centerline{{\bf Shyamoli Chaudhuri and Joseph D. Lykken}\footnote{$^\dagger$}
{e-mail:\ shyam@FNAL, and lykken@FNAL}}
\bigskip\centerline{Theory Group, MS106}
\centerline{Fermi National Accelerator Laboratory}
\centerline{P.O. Box 500, Batavia, IL 60510}


\vskip .3in
We analyse in detail the $SL(2,R)$ black hole by extending standard
techniques of Kac-Moody current algebra to the non-compact case. We construct
the elements of the ground ring and exhibit $W_{\infty}$ type structure in the
fusion algebra of the discrete states. As a consequence, we can identify
some of the exactly marginal deformations of the black hole. We show that
these deformations alter not only the spacetime metric but also turn on
non-trivial backgrounds for the tachyon and {\it all} of the massive modes of
the string.

\Date{6/92} 

%
\def\cmp#1{{\it Comm. Math. Phys.} {\bf #1}}

\def\plb#1{{\it Phys. Lett.} {\bf #1B}}
\def\pl#1{{\it Phys. Lett. } {\bf #1B}}

\def\prd#1{{\it Phys. Rev.} {\bf D#1}}

\def\npb#1{{\it Nucl. Phys.} {\bf B#1}}

\def\mpl#1{{\it Mod. Phys. Lett.} {\bf A#1}}

\lref\ginsparg{P. Ginsparg and F. Quevedo, ``Strings on curved
spacetimes: black holes, torsion, and duality'', Los Alamos
preprint LA-UR-92-640, February 1992.}
\lref\witd{E. Witten, ``Two-dimensional string theory and
black holes'', Inst. for Adv. Study preprint
IASSNS-HEP-92-25, April 1992.}
\lref\louck{L. C. Biedenharn and J. Louck, {\it Ann. Phys.} {\bf 63}
(1971) 459.}
\lref\polc{J. Polchinski, \npb{346} (1990) 253.}
\lref\callan{C. Callan, S. Giddings, J.A. Harvey, and
A. Strominger, \prd{45} (1992) 1005.}
\lref\lenny{J. Russo, L. Susskind and L. Thorlacius, ``Black hole evaporation
in 1+1 dimensions'', SU-ITP-92-4, January 1992; ``The end point of black hole
evaporation'', SU-ITP-92-17, June 1992.  L. Susskind and L. Thorlacius,
``Hawking radiation and back reaction'', SU-ITP-92-12, April 1992.}
\lref\banks{T. Banks, A. Dabholkar, M.R. Douglas, and M. O'Loughlin,
\prd{45} (1992) 3607.}
\lref\bfnpap{C. Callan, D. Friedan, E. Martinec and M. Perry, \npb{262} (1985)
593.}
\lref\bars{I. Bars and D. Nemeschansky, \npb{348} (1991) 89.  I. Bars and
K. Sfetsos, \pl{277} (1992) 626.}
\lref\sbguys{J. Horne and G.T. Horowitz, \npb{368} (1992) 444.}
\lref\kuta{M. Bershadsky and D. Kutasov, \plb{266} (1991) 345.}
\lref\raiten{E. Raiten, ``Perturbations of a stringy black hole'', Fermilab
preprint Fermi-91/338-T, December 1991.}
\lref\mandal{G. Mandal, A. Sengupta and S. Wadia, \mpl{6} (1991) 1685.}
\lref\rabi{S. Elitzur, A. Forge and E. Rabinovici, \npb{359} (1991) 581.}
\lref\morse{P. Morse and H. Feshbach, {\it Methods of Theoretical Physics},
McGraw Hill, 1953.}
\lref\distler{J.Distler and P. Nelson, \npb{366} (1991) 255.}
\lref\wit{E. Witten, \prd{44} (1991) 314.}
\lref\dlg{N. Sakai and I. Senda, Prog. Theor. Phys. {\bf75} (1986) 692.}
\lref\kz{V. G. Knizhnik and A. B. Zamolodchikov, \npb{247} (1984) 83.}
\lref\gad{K. Gawedzki and A. Kupianen, \pl{215} (1988) 119. D. Karabali,
Q.-H. Park, H. Schnitzer and Z. Yang, \pl{216} (1989) 307.
K. Gawedzki, Cargese lectures, 1991.}
\lref\dealwis{S. DeAlwis and J. Lykken, \pl{269} (1991) 264.}
\lref\dpl{L. Dixon, J. Lykken and M. Peskin, \npb{325} (1989) 329.}
\lref\god{P. Goddard, A. Kent and D. Olive, \pl{152} (1985) 88; \cmp{103}
(1986) 105.}
\lref\hwa{M. Henningson, S. Hwang, P. Roberts, and B. Sundborg,
\pl{267} (1991) 350.}
\lref\pet{P. Petropoulos, \pl{236} (1990) 151.}
\lref\barg{V. Bargmann, {\it Ann. Math.} {\bf 48} (1947) 568.}
\lref\meb{J. Lykken, \npb{313} (1989) 473.}
\lref\grif{P. Griffin and O. Hernandez, \npb{356} (1991) 287.}
\lref\gep{D. Gepner and E. Witten, \npb{278} (1986) 493.}
\lref\verlinde{R. Dijkgraaf, H. Verlinde, and E. Verlinde,
``String propagation in a black hole geometry'', Princeton
preprint PUPT-1252, May, 1991.}
\lref\ground{E. Witten, \npb{373} (1992) 187.}
\lref\vilenkin{N. J. Vilenkin, {\it Special Functions and the Theory of Group
Representations}, AMS, 1968.}
\lref\bieden{W. I. Holman and L. C. Biedenharn, Jr., Ann. of Phys. {\bf 34}
(1966) 1.}
\lref\seiberg{G. Moore and N. Seiberg, \npb{362} (1991) 665.
N. Seiberg and S. Shenker, ``A note on background (in)dependence'',
Rutgers preprint RU-91-53, January 1992.}
\lref\bpz{A. A. Belavin, A. M. Polyakov, and A. B. Zamalodchikov, \npb{241}
(1984) 333.}
\lref\barton{E. Witten and B. Zweibach, ``Algebraic structure and differential
geometry in 2-d string theory'', IAS preprint IASSNS-HEP-92-4, January 1992.}
\lref\kadanoff{L. P. Kadanoff and A. C. Brown, Ann. Phys. 121 (1979) 318.}
\lref\radius{P. Ginsparg, \npb{295} [FS21] (1988) 153.
R. Dijkgraaf, E. Verlinde
and H. Verlinde, \cmp{115} (1988) 649.}
\lref\chsch{S. Chaudhuri and J. A. Schwartz, \pl{219} (1989) 291.}
\lref\abrom{M. Abromowitz and I. Stegun,
{\it Handbook of Mathematical Functions},
Dover, 1964.}
\lref\bardacki{K. Bardacki, M. Crescimanno, and E. Rabinovici, \npb{344}
(1990) 344.}
\lref\kleb{I. Klebanov and A. M. Polyakov, \mpl{6} (1991) 3273.}
\lref\szeg{E. Bergshoeff, C. N. Pope, L. J. Romans, E. Szegin and X. Shen,
\pl{245} (1990) 447.}
\lref\pope{C. N. Pope, L. J. Romans and X. Shen, \pl{136} (1990) 173.}
\lref\shen{X. Shen, ``W-infinity and string theory'', CERN preprint
CERN-TH-6404-92, February 1992.}
\lref\ellis{J. Ellis, N. E. Mavromatos and D. V. Nanopoulos, \pl{276} (1992)
56; CERN-TH-6413-92,-6514-92 and 6476-92.}
\lref\barbon{J. L. F. Barbon, ``Perturbing the ground ring of 2-d string
theory'', CERN preprint CERN-TH-6379-92, January 1992.}
\lref\dot{V. Dotsenko, ``The operator algebra of the discrete state operators
in 2-d gravity with non-vanishing cosmological constant'', CERN preprint
CERN-TH-6502-92, January 1992.}
\lref\romans{L. Romans, \npb{357} (1991) 549.}
\lref\ber{M. Bershadsky, \cmp{139} (1991) 71.}
\lref\ritten{V. Rittenberg and M. Scheunert, \cmp{83} (1982) 1.}
\lref\popeshen{C. N. Pope and X.Shen, \plb{236} (1990) 21.}
\lref\popeb{C. N. Pope, L. J. Romans and X. Shen, \pl{242} (1990) 401.}
\lref\kir{I. Bakas and E. Kiritsis, ``Beyond the large N limit:
non-linear $W_\infty$ as symmetry of the $SL(2,R)/U(1)$ coset
model'', UCB-PTH-91-44, September 1991.}
\lref\puk{L. Pukanzsky, ``On the Kronecker products of irreducible
representations of the $2$$\times$$2$ real
unimodular group'', {\it Trans. Am. Math. Soc.} {\bf 100} (1961) 116.}


\nfig\camodule{Current algebra representation built on a discrete series
representation of $SL(2,R)$. The number in each circle indicates the
multiplicity of states at that value of $(L_0, J^3_0)$.}

%
\def\nl{&\cr\tablerule}
\newbox\bigbox
\setbox\bigbox=\hbox{\vrule height15pt depth8pt width0pt}
\def\bigstrut{\relax\ifmmode\copy\bigbox\else\unhcopy\bigbox\fi}
\def\sp{{s^\prime}}
\def\pt{\tilde\Phi}
\def\vt{\tilde V}
\def\gt{\tilde g}
\def\at{\tilde a}
\def\bt{\tilde b}
\def\lt{\tilde L}
\def\GP#1{G^{({\sss +})#1}}
\def\GM#1{G^{({\sss -})#1}}
\def\GPM#1{G^{({\sss\pm})#1}}
\def\sss{\scriptscriptstyle}
\def\slr{$SL(2,R)$}

\def\gb{{\bf g}}
\def\gp{g_{{\sss ++}}}
\def\gm{g_{{\sss --}}}
\def\fm{g_{{\sss -+}}}
\def\fp{g_{{\sss +-}}}
\def\gpp{g_{{\sss\pm\pm}}}
\def\gpm{g_{{\sss\pm\mp}}}
\def\gmp{g_{{\sss\mp\pm}}}
\def\j#1{J^{{\sss #1}}}
\def\jb#1{\bar J^{{\sss #1}}}
\def\jt{J^{{\scriptstyle 3}}}
\def\wwp{W^{{\sss +}}}
\def\wp{W^{hw}}
\def\wm{W^{lw}}
\def\wc{W^{cont}}

\def\pp{\Phi ^{hw}}
\def\pmi{\Phi ^{lw}}
\def\pc{\Phi ^{cont}}
\def\ppp{\Phi ^{hw-hw}}
\def\pmm{\Phi ^{lw-lw}}
\def\ppm{\Phi ^{hw-lw}}
\def\pmp{\Phi ^{lw-hw}}
\def\op{O^{hw}}
\def\om{O^{lw}}
\def\oc{O^{cont}}
\def\od{O^{double}}

\def\co{{\rm cosh}\,}
\def\si{{\rm sinh}\,}
\def\ta{{\rm tanh}\,}
\def\se{{\rm sech}\,}
\def\cs{{\rm csch}\,}
\def\e{{\rm exp}\,}
\def\lo{{\rm ln}\,}
\def\js{{\bf J}^2}
\newsec{Introduction}

String theory is widely advertised as the only viable candidate
for a complete, consistent theory of quantum gravity coupled to matter.
As such, it must give unambiguous answers to the plethora of
important questions left unresolved by traditional semiclassical
treatments of gravitating systems. Indeed, this is the arena in which
string theory {\it must} deliver profound insights, or else give
up its claim to be the fundamental theory of Planck scale physics.
Many of the most interesting and important questions in
quantum gravity concern the physics of black holes. These questions
involve quantum decoherence, Hawking radiation, the endpoint of black
hole evaporation, the existence of naked singularities, and the
quantum numbers (``hair'') of black hole solutions.

Some insight into black hole physics has already been obtained
from general string arguments as well as string-inspired
methods\callan ,\lenny ,\banks ,\ellis . At the same time,
there has been substantial progress in developing new and better
techniques to handle strings in nontrivial spacetime
backgrounds, beyond the beta function methods introduced some
years ago by Callan et al\bfnpap . At least in two spacetime
dimensions, matrix model techniques seem to offer hope of obtaining
rigorous nonperturbative results for strings in nontrivial
backgrounds. Collective field and string field theory approaches
also have great promise in the long term.
Another strategy is to develop conformal field theories that
describe strings in nontrivial spacetime backgrounds. This
approach is less rigorous and complete than matrix models
or string field theory, but has the advantages of being somewhat
more familiar and closer to the physics. On the other hand,
this approach is more rigorous and complete than beta function
or string-inspired methods, but has the disadvantage of being
more abstract and unwieldy.

In \wit , Witten suggested that a gauged $SL(2,R)$ Wess-Zumino-Witten
(WZW) sigma model can provide a conformal field theory model of
a two dimensional bosonic string in a black hole background.
This model, as well as variations of it,
was further analyzed in \verlinde ,\kuta ,\distler ,
and many other papers. It should be emphasized that the
conformal field theory approach to black holes is in no way
limited to two dimensions; indeed extensions of Witten's model
to three and four dimensional black holes have already been
exhibited\bars ,\sbguys ,\raiten ,\ginsparg .

In this paper we will analyze the $SL(2,R)$ black hole in
more detail, by extending standard techniques of Kac-Moody
current algebra. While we have not resolved all of the sticky
issues which appear in noncompact coset models, we are able
to pin down the physical content of the theory to a considerable
extent. We show that the discrete state operators, previously
exhibited by Distler and Nelson, form a $W_\infty$ type algebra,
and may thus be regarded as generating infinite quantum
``$W$ hair'' of the stringy black hole. From this structure we
are able to identify some of the exactly marginal deformations
of the black hole. These deformations not only alter the
spacetime metric, but also turn on nontrivial backgrounds
for the tachyon (which is really a massless scalar) and
{\it all the massive modes of the string}.

Our results appear to confirm previous speculation\ellis \witd\
concerning the fundamental interplay of string physics
with black hole physics. We also make contact with previous
results from beta function
calculations\mandal ,\polc ,\dealwis , the ground
ring approach of Witten\ground , as well as results from
two dimensional Liouville theory.

\newsec{The $SL(2,R)$ String}

Following Witten\wit\ we consider an \slr\ WZW sigma
model with action

\eqn\wzwaction{
{k\over 8\pi}\int d^2z\;\sqrt{h}h^{ij}\;Tr\left(
\gb ^{-1}\del _i \gb\;\gb ^{-1}\del _j \gb\right) \; +ik\Gamma\ ,
}
where $h$ is the worldsheet metric, $\Gamma$ is the Wess-Zumino term,
and the field $\gb (z,\bar z)$ is an element of \slr . This action is
invariant (up to a surface term) under global $SL(2,R)_L\times SL(2,R)_R$
transformations.
It is convenient to parametrize the components of $\gb$ as follows

\eqn\compdef{
\gpp =Tr\left( R_{\pm}\gb \right)\, ,\quad \gpm =
Tr\left( S_{\pm}\gb \right) ,
}
where $R_\pm$ and $S_\pm$ are defined in terms of the unit matrix
and the Pauli matrices:

$$
R_\pm = \half\left( 1\pm\sigma _2\right)\, ,\quad
S_\pm = \half\left( \sigma _1\mp i\sigma _3\right) .$$

\noindent The components $\gp$, $\gm$, $\fp$, and $\fm$ also parametrize
$SU(1,1)$ (which is isomorphic to \slr ) and in fact transform
as four spin $(\half ,\half )$ components under $SU(1,1)_L\times
SU(1,1)_R$. Note that $\gm$$=$$\gp ^*$, \ $\fp$$=$$\fm ^*$, and
we have the constraint

\eqn\constraint{
\gp\gm - \fp\fm = 1 .
}

While the parametrization above is convenient for current algebra,
it is better when discussing target space physics to use Euler angles:

\eqn\euler{
\gb (z,\bar z) = e^{\half i \theta_L \sigma_2 } e^{r \sigma_1}
e^{\half i \theta_R \sigma_2} .
}

\noindent Translation between the two parametrizations is given by

\eqn\translate{
\gpp =\co r\,\e \pm\half i(\theta _L + \theta _R)\, ,\quad
\gpm =\si r\,\e \mp\half i(\theta _L - \theta_R) .
}

As in \wit , we introduce an abelian gauge field $A(z,\bar z)$ to gauge
the axial diagonal $U(1)$ subgroup of $SL(2,R)_L\times SL(2,R)_R$.
If we then fix to unitary gauge: $\theta _L$$=$$-\theta _R$$=$$\theta$,
and integrate out the gauge field $A$, the action reduces to the Euclidean
black hole background

\eqn\bhb{
{k\over 4\pi}\int d^2z\, \left( \del _zr\del _{\bar z}r +
{\rm tanh}^2r\, \del _z\theta \del _{\bar z}\theta \right) .
}

\noindent Also, by evaluating the
determinant that arises in integrating out the gauge
field, one obtains (to lowest order) the dilaton term\dlg

\eqn\dlt{
\Phi = 2\, \lo \co r .
}

For $k$$=$$9/4$, this gauged WZW model appears to be a conformal field
theory description of a bosonic string embedded in a two dimensional
target space with a black hole background metric.
There are difficulties involved in the consistent quantization of
gauged noncompact WZW models\gad , but in this paper we will simply assume
that Witten's \slr\
model exists as a consistent unitary conformal field theory. Our purpose
will be to analyze the content of this theory by employing \slr\ (or rather,
equivalently, $SU(1,1)$) Kac-Moody current algebra.

The holomorphic and antiholomorphic $SU(1,1)$ currents are the
composite operators\kz

\eqn\currents{
\eqalign{ \j{\pm} (z) =
&\kappa ~(~ \gpm ~ \partial_z \gpp - \gpp ~
\partial_z \gpm ~ ) ,\cr
J^{3} (z) = &\kappa ~(~ \gp ~ \partial_z \gm +
\fp ~ \partial_z \fm ~ ) ,
\cr
\jb{\pm} (\bar z ) =
&\kappa ~(~ \gmp ~ \partial_{\bar z} \gpp - \gpp ~
\partial_{\bar z} \gmp ~) ,\cr
\bar J^{3} (\bar z) = &\kappa ~(~ \gm ~ \partial_{\bar z} \gp +
\fm ~ \partial_{\bar z} \fp ~ ) ,
\cr
}
}
\noindent where $\kappa = k$$-$$2$$=$$1/4$. Note that,
here and below, normal ordering is implied in operator composites.
The operator product expansions of these
composites with $\gb$ are given by \kz\

\eqn\ope{\eqalign{
J^a(z) \gb (w, \bar w) &= {{\gb \tau^a}\over{z-w}}
+ : J^a(z) \gb (w, \bar w):~ , \cr
\bar J^a (\bar z) \gb (w, \bar w ) &= {{\tau^a \gb }\over{\bar z - \bar w }} +
: \bar J^a (\bar z ) \gb (w, \bar w ): ~, \cr}
}
where $\tau^1 = i\sigma_1 /2$,
$\tau^2=\sigma_2 /2$, $\tau^3=i\sigma_3 /2$, are the generators of $SU(1,1)$.

Each set of currents obeys an $SU(1,1)$ current algebra. For example,
the conformal modes of the holomorphic currents satisfy\dpl

\eqn\kma{
\eqalign{ \left[ J^3_n , J^3_m \right] &= -\half kn\, \delta _{n+m,0}
,\cr
\left[ J^3_n , \j{\pm}_m \right] &= \pm \j{\pm}_{n+m}
,\cr
\left[ \j{+}_n , \j{-}_m \right] &= kn\, \delta _{n+m,0} - 2J^3_{n+m} .\cr
}
}
The Virasoro generators are given by the Sugawara construction:

\eqn\stress{
T(z) = {1\over\kappa}J^i(z)J^i(z)
,}
with $c=3k/\kappa =27$.
Acting on states in a Kac-Moody module,
$L_0$ is simply related
to the $SU(1,1)$ zero-mode Casimir:

\eqn\lzero{
L_0 = {1\over\kappa}\js  + N
,}
where $N$ is a nonnegative integer called the {\it grade} of a state in
a Kac-Moody module, while the zero-mode Casimir is given by

\eqn\casimir{
\eqalign{
\js &= \half \left( \j{+}_0 \j{-}_0 + \j{-}_0 \j{+}_0 \right)
- \left( J^3_0 \right)^2
,\cr
&= -j(j+1) .\cr
}
}

Gauging the diagonal $U(1)$ in the WZW model is equivalent to modding
out by all $U(1)$ descendant states, both holomorphically and
antiholomorphically, in every Kac-Moody module. The
$SU(1,1)/U(1)$ coset has $c=26$.
Acting on states in a coset module, $L_0$ has eigenvalues

\eqn\newlzero{
L_0 = -4j(j+1) + {4\over 9}m^2 + N
,}
where $m$ is the $J^3_0$ eigenvalue of a state in the coset module.

Although coset
conformal field theories based on compact groups, e.g. $SU(2)/U(1)$, are
very well understood\god , noncompact cosets are poorly understood
and are not even known to exist as unitary modular invariant theories
except in special cases\dpl ,\hwa ,\pet .
Furthermore, for purposes of doing string theory with $SU(1,1)/U(1)$
cosets, we must be very general about introducing all Kac-Moody coset
modules which contain string physical states, i.e. coset states which
obey the Virasoro highest weight and mass-shell conditions

\eqn\mshell{
\eqalign{
L_n\, | j,m,N > &= 0\; ,\quad {\rm for\ all\ }n>0, \cr
\left( L_0 -1 \right) | j,m,N > &= 0  ,\cr
}
}
and thus correspond to dimension $(1,1)$ operators in the coset theory.
While we should require that the totality of such operators form a
closed local operator algebra, and
that the corresponding physical states
have all relatively nonnegative norms, this in no way implies that
we can restrict our attention to unitary $SU(1,1)/U(1)$ modules.
In fact it turns out\distler\ that to obtain all the physical
states one needs to consider even Kac-Moody modules which have
negative norm states at the base (i.e. grade zero)!

With this in mind we now briefly review the representation theory
of classical $SU(1,1)$\barg\ and of $SU(1,1)/U(1)$ Kac-Moody\dpl .
The $SU(1,1)$ Kac-Moody modules can be characterized by the states
which occur at the base, the Kac-Moody primaries. These states are
annihilated by all the $\j{\pm}_n$ and $\jt _n$ for all $n>0$, and
together constitute a module of the zero-mode $SU(1,1)$. Thus the
Kac-Moody primaries correspond to representations (reps) of classical
$SU(1,1)$. As shown by Bargmann, the irreducible representations (irreps) of
classical $SU(1,1)$ are either {\it double-sided}, {\it highest weight
discrete series}, {\it lowest weight discrete series}, or
{\it continuous series}. The double-sided representations are the
only finite dimensional reps of $SU(1,1)$, and are isomorphic to the
standard unitary irreps of $SU(2)$. However for $SU(1,1)$ these reps are
all nonunitary, excepting only the trivial identity representation.

The highest weight and lowest weight discrete series irreps
contain a state annihilated by $\j{+}_0$ or $\j{-}_0$, respectively.
They are thus one-sided and infinite dimensional. A highest weight
state has $m=j$ or $m=-j$$-$$1$, while a lowest weight state has
$m=-j$ or $m=j$$+$$1$. If we apply no constraints from unitarity
or single-valuedness, then $j$ can be any real number, although
in the cases where $j$ is a nonnegative integer or half-integer
the corresponding highest and lowest weights reps degenerate to the
double-sided reps. In addition, because of the equivalence of irreps
related by $j\rightarrow -j$$-$$1$, we can restrict $j$ to values
either $\ge$ or $\le$\ $-1/2$. In this paper we will restrict to
negative $j$ values

\eqn\jvalues{
j\le -1/2 ,
}
\noindent for all discrete series representations. With this convention
we may observe that the $m=j$ highest weight and $m=-j$ lowest weight
irreps are unitary, while the $m=\mp (j$$+$$1)$ irreps are nonunitary.

The continuous series irreps of classical $SU(1,1)$ have neither a highest
nor a lowest weight state, which merely requires that
neither $j$$+$$m$ nor $j$$-$$m$ is an integer. While $m$ is real, $j$ can
be complex. In fact the unitary continuous series reps have
either $j=\, -1/2$$+$$i\rho$ or $-1/2<j<0$, i.e. $\js >0$.
However, for
our purposes we will need the nonunitary continuous series
reps with $\js <1/4$. Indeed it will suffice to consider only
continuous series reps which have $j$ real and restricted to \jvalues .

Given a set of Kac-Moody primaries which form an irrep of the zero-mode
$SU(1,1)$, the Kac-Moody module is built up by applying the raising
operators $\j{\pm}_{-n}$ and $\jt_{-n}$. The number of independent states
with a particular $m$ value at a particular grade in any module
(excepting the special cases described below) is given
by counting the number of ways such states can be obtained by
the free action of the raising operators- i.e., by
acting with
strings of raising operators on all states in the base, modulo strings which
differ only in their ordering. For example in Fig. 1 we show the
multiplicities of a generic lowest weight module.

The reason why the multiplicities of the discrete and continuous series
$SU(1,1)$ Kac-Moody modules are so simply determined is that {\it these
modules generically contain no nontrivial null states}. This is easily
seen by applying the ``pseudospin'' analysis used by Gepner and Witten\gep
to study affine $SU(2)$. Like $SU(2)$, affine $SU(1,1)$ has an external
automorphism symmetry corresponding to permuting the weights of the
extended Dynkin diagram. As a result, an arbitrary module can be decomposed
either into reps of the zero-mode $SU(1,1)$ or into reps of the
$SU(1,1)$ pseudospin defined by

\eqn\pseudo{
\eqalign{
\left[ \hat J^3_0\, , \,\j{+}_1 \right] &= \j{+}_1 ,\cr
\left[ \hat J^3_0 , \j{-}_{-1} \right] &= -\j{-}_{-1} ,\cr
\left[ \j{+}_1 , \j{-}_{-1} \right] &= -2 \hat J^3_0 ,\cr
}
}
\noindent where

\eqn\newj{
\hat J^3_0 \equiv \jt _0 - \half k .
}
Now consider, for example, an $m=-j$ lowest weight Kac-Moody primary
of a lowest weight module. This state is obviously also a highest
weight state with respect to pseudospin, with $\hat m=-j$$-$$k/2$.
Thus, excepting the special cases where $\hat m$ is a nonnegative integer
or half-integer, this lowest weight state also defines a highest weight
irrep of pseudospin, which contains all of the states on the diagonal
boundary of the module (see Fig. 1). It follows from the automorphism symmetry
that every state in the module belongs to both a lowest weight irrep of the
zero-mode algebra and a highest weight irrep of the pseudospin algebra.
Thus the null state structure is trivial. It is amusing to note that,
for $k=9/4$, the double-sided $SU(1,1)$ modules will look nothing like
affine $SU(2)$ modules, even though they have an $SU(2)$-like
representation at the base. This is because for $k=9/4$ each state at
the base of a double-sided module necessarily generates a highest weight
irrep of $SU(1,1)$ pseudospin, not a double-sided one.

$SU(1,1)/U(1)$ coset modules are obtained from $SU(1,1)$ modules by
modding out all $U(1)$ descendant states, i.e. keeping only states
which satisfy the $U(1)$ highest weight condition:

\eqn\hwcond{
\jt _n\, |j,m,N > = 0\; ,\quad {\rm for\ all\ }n>0 .
}
\noindent The coset modules can also be described using
parafermions\dpl ,\meb ,\grif .

\newsec{Physical States of the $SL(2,R)$ String}

Using BRST cohomology,
Distler and Nelson\distler\ have done a complete holomorphic
classification of the maximal set of independent physical states
in the $SL(2,R)$ coset string theory (the actual physical states
may be some truncation of this set).
They are of three types: tachyon
states, discrete states, and Virasoro null states. We will defer
discussion of the Virasoro null states until the next section.
The tachyon states consist of
Kac-Moody primaries of dimension 1; they have arbitrary
real $j\le -1/2$, and have $m=\,\pm(3j+3/2)$. These states appear to
correspond to the normalizable tachyon states of the $c$$=$$1$
Liouville theory\distler , with $j$ playing the role of the Liouville
momentum.

The discrete states are physical states appearing at higher grade
in certain $SU(1,1)$ Kac-Moody modules for discrete values of $j$.
Their $(j,m)$ quantum numbers can be parametrized by two positive
integers $s$ and $r$. There are discrete states occurring in highest
weight modules which have unitary ($m$$=$$j$ type) representations at the
base. Their quantum numbers are:

\eqn\hwphys{
j= -{1\over 4}(s+2r+1)\; ,\quad m={3\over 4}(s-2r+1) .
}

\noindent There are discrete states occuring in lowest
weight modules which have unitary ($m$$=$$\,-j$ type) representations at the
base. Their quantum numbers are:

\eqn\lwphys{
j= -{1\over 4}(s+2r+1)\; ,\quad m=-{3\over 4}(s-2r+1) .
}
In addition, there are discrete states appearing in continuous
series modules which have nonunitary irreps at the base:

\eqn\cphys{
j= -{1\over 2}(s+r+1)\; ,\quad m={3\over 2}(s-r) .
}
\noindent For each of these states, the grade is trivially computed
from \mshell .

We should note that the discrete states listed above are only half of
the states given in \distler . However the remaining
discrete states map into the
states above under the external automorphism. Our attitude will be to ignore
this duplication, as well as the probably infinite cloning of states due
to nontrivial winding sectors. Some insight on these issues can be found
in \hwa ,\verlinde ,\distler .

As is done for the Liouville theory, it will be useful to augment the
discrete states by a certain subset of the tachyon states. This is
because the operator algebra of the discrete states cannot close
on itself without the addition of these ``discrete'' tachyon states.
With our conventions these additional states are the dimension 1
Kac-Moody primaries which have $j=\,-r/4$, $r=2,3,4,\ldots$.

With the above caveats and additions, Table 1 gives a listing of
the first 34 discrete states, paired with Liouville states
in the manner suggested in \distler .
The notation $\wwp _{s,n}$ for Liouville states is as in \ground .
Our notation for $SL(2,R)$ holomorphic discrete states is $\wp _{j,m}$
for states in highest weight modules, $\wm _{j,m}$ for states in lowest
weight modules, and $\wc _{j,m}$ for states in continuous series modules.
Also we denote Kac-Moody primaries by $\pp _{j,m}$, $\pmi _{j,m}$, and
$\pc _{j,m}$. This table should not be taken very seriously. As pointed
out in \distler , the correspondence suggested between Liouville and
$SL(2,R)$ is not one-to-one, due to ``extra'' $SL(2,R)$ discrete
states. We will see later, when we discuss the algebra of $SL(2,R)$
discrete state operators, that the correspondence to Liouville is in fact
quite a bit more complicated than suggested by Table 1.

So far, the discussion of physical states has been purely holomorphic.
However in the WZW model physical state operators are composites of
the components of $\gb (z,\bar z)$
(and $A$ in the gauged version) which are not
holomorphic fields. Holomorphic physical states as described above do not
actually exist in this theory. We can, however, explicitly construct
operators as composites of $\gpp (z,\bar z)$ and $\gpm (z,\bar z)$
whose holomorphic content matches that above. We find from this
explicit construction that the rules for tieing together holomorphic
and antiholomorphic sectors in the $SL(2,R)$ WZW model are rather
simple. Modulo questions of normalizability and single-valuedness,
the only constraint is $j=\bar\jmath$. Furthermore, modules of different
type but the same $j$ can be tied together. Thus for example a highest
weight module with fixed $j$ can tie up with another highest weight
module, or a lowest weight, or a continuous series, or a doubled-sided
module (for suitable $j$).

To be more specific, we have already given explicit expressions for
the WZW currents in \currents . One can then construct any state in
the theory using $SL(2,R)_L\times SL(2,R)_R$ current algebra,
provided one has an explicit construction of the Kac-Moody
primaries. The action of the
zero mode currents on the spin half primaries \compdef\ is given by the
singular terms in \ope , and their descendants are
obtained from the Taylor expansion of the second (regular) term.
Using Wick's theorem we can extend this construction to arbitrary primaries,
some examples of which are given in Section 6.
In the appendix we derive a general expression for
an arbitrary Kac-Moody primary in the (ungauged) $SL(2,R)$ WZW
theory:

\eqn\hyper{
\eqalign{
\Phi _{j,m,\bar m}(z&,\bar z) = \left[ \Gamma (j+m+1) \Gamma (j-m+1)
\Gamma (j+\bar m+1) \Gamma (j-\bar m+1) \right]^{1/2}
\cr
&\times \sum ^\infty _{n=-\infty}
{
(\gp )^{j+m}(\gm )^{j-\bar m}(\fm )^{\bar m-m}
(-\fp\fm /\gp\gm )^n \over
\Gamma (j+m-n+1)\Gamma (j-\bar m-n+1)\Gamma (\bar m-m+n+1)
\Gamma (n+1)} .\cr
}
}
\noindent For primaries which transform like
continuous-continuous under $SL(2,R)_L\times SL(2,R)_R$, this expression
reduces to a hypergeometric function \vilenkin, \verlinde . For example:

\eqn\ccexp{
\Phi ^{c-c}_{-1/2,0,0}(z,\bar z) = (\gp\gm )^{-1/2} F(\half ,\half ;1;
\fp\fm /\gp\gm ) ,
}

\noindent which is equivalent to the integral expression given in
\verlinde . Note $\Phi ^{c-c}_{-1/2,0,0}$ is (roughly speaking)
the $SL(2,R)$ analog of the cosmological constant
operator in the Liouville theory.

For other primaries \hyper\ reduces to a simple expression.
For example:

\eqn\jzero{
\ppp _{-1,-1,-1}(z,\bar z) = {1\over\gp ^2}\, ,\quad
\pmm _{-1,1,1}(z,\bar z) = {1\over\gm ^2} .
}

We should warn the reader that it is not clear whether all of the
formal
composites implied by \hyper\ really exist as well-defined operators
creating normalizable states (or even non-normalizable states in the
sense of \seiberg ).
The normalization convention chosen in \hyper\
is reasonable for $\ppp$, $\pmm$, $\ppm$, and $\pmp$ states, but
certainly not for more exotic states.
Some of the physical states are definitely not square integrable with
respect to the classical $SL(2,R)$ invariant measure. For example
\ccexp\ is not square integrable,
though the rest of the continuous-continuous physical
states are. The identity is also not square integrable\bieden .

\newsec{The Ground Ring}

In addition to the physical states already discussed, the $SL(2,R)/U(1)$
string theory has an infinite number of physical Virasoro null states.
These correspond to physical states of ghost number zero in
BRST language, and are the analogues of the ``ground ring'' operators
introduced in \ground .
The existence of these null states is a direct result of a ``pathological''
property of the $k$$=$$9/4$\ $SU(1,1)/U(1)$ Kac-Moody coset modules, namely,
the presence of an infinite number of operators with conformal dimension
zero.
By contrast, in a typical conformal field theory only the identity operator
has zero dimension. These dimension zero operators are
a result of the fact that the zero-mode $SU(1,1)$ Casimir makes a negative
contribution to the mass operator \newlzero\ for $j$$<$$-1$.
Liouville theory, of course, has similar properties.

Let us consider a coset state which is Virasoro highest weight
and has dimension zero. Then
the coset Virasoro algebra implies that $L_{-1}$ of this state
is simultaneously Virasoro highest weight and a Virasoro
descendant. Thus it either vanishes identically,
or it is a Virasoro null state. Since $L_{-1}O=\del _zO$ when
$O$ is Virasoro primary, $L_{-1}$ of the identity is the only case
in which null states of this construction vanish identically.

The $(j,m)$ quantum numbers
of the independent dimension zero physical states are given in \distler .
In Table 2 we list the first 12 dimension zero coset operators.
Our notation is
$\op _{j,m}$, $\om _{j,m}$, $\oc _{j,m}$, and $\od _{j,m}$ for
operators in highest weight, lowest weight, continuous series, and
double-sided modules.
We have grouped the operators by their $j$ values, and listed in addition
on the same line all of the non-null physical states with the same $j$.
Up to questions of normalizability, single-valuedness,
and closure, the $k$$=$$9/4$
gauged $SL(2,R)$ WZW theory will contain operators that behave
like any listed operator of given $j$ holomorphically and any other
listed operator of the same $j$ antiholomorphically. Of particular
interest are the dimension $(1,0)$ and $(0,1)$ operators of the form:

\eqn\grcurrents{
J_{j,m,\bar m}(z,\bar z) \equiv W_{j,m}\bar O_{j,\bar m}\; ,\quad
\bar J_{j,m,\bar m}(z,\bar z) \equiv O_{j,m}W_{j,\bar m} .
}

\noindent As discussed in \ground ,\barton\ for
the Liouville theory, such operators are purely
holomorphic/antiholomorphic up to string null states. They may
therefore be regarded as an infinite set of conserved physical
currents, with corresponding conserved charges.

In the Liouville theory there is a chiral ``ground ring'' of dimension
zero operators $O^{\sss +}_{j,m}$ generated by
$X=O^{\sss +}_{1/2,1/2}$
and $Y=O^{\sss +}_{1/2,-1/2}$. A somewhat analogous structure appears
for the string in a black hole background.
We may consider the dimension zero
coset Virasoro primaries

\eqn\xandy{
X = O^{lw}_{-3/4,3/4}\; ,\quad Y = O^{hw}_{-3/4,3/4} .
}
These operators correspond to grade 1 states built up from the
lowest/highest weight Kac-Moody primaries $\Phi ^{lw}_{-3/4,-1/4}$
and $\Phi ^{hw}_{-3/4,1/4}$, and thus live in modules which are
nonunitary even at the base. They have the explicit form

\eqn\xyexp{
\eqalign{
X &= 2\sqrt{{2\over 11}}\left[ \j{+}_{-1} + {16\over 3}\jt _{-1}
\j{+}_0 - 5\j{-}_{-1}\j{+}_0\j{+}_0 \right] \Phi ^{lw}_{-3/4,-1/4}
,\cr
Y &= 2\sqrt{{2\over 11}}\left[ \j{-}_{-1} + {16\over 3}\jt _{-1}
\j{-}_0 - 5\j{+}_{-1}\j{-}_0\j{-}_0 \right] \Phi ^{hw}_{-3/4,1/4} .\cr
}
}

\noindent The corresponding null states are obtained by applying
$L^{coset}_{-1}$, which we can
effectively write in terms of the $SU(1,1)$
Virasoro operators and currents as

\eqn\nullguy{
L^{coset}_{-1}X = \left[ L_{-1} + {3\over 2k}J^3_{-1} \right] X .}
Actually $X$ and $Y$
should be regarded as shorthand for the holomorphic
content of the four operators

\eqn\aas{
a_1(z,\bar z)=X\bar X\, ,\; a_2(z,\bar z)=Y\bar Y\, ,\;
a_3(z,\bar z)=X\bar Y\, ,\; a_4(z,\bar z)=Y\bar X
,}

\noindent which are built up, respectively, from the four
Kac-Moody primaries

\eqn\fourkm{
\eqalign{
\Phi ^{lw-lw}_{-3/4,-1/4,-1/4}(z,\bar z) &=
\sqrt{\gm}\; ,\quad
\Phi ^{hw-hw}_{-3/4,1/4,1/4}(z,\bar z) =
\sqrt{\gp} ,\cr
\Phi ^{lw-hw}_{-3/4,-1/4,1/4}(z,\bar z) &=
\sqrt{\fm}\; ,\quad
\Phi ^{hw-lw}_{-3/4,1/4,-1/4}(z,\bar z) =
\sqrt{\fp} .\cr
}
}

Let us now consider the operator product of $X$ and $Y$. As in
\ground , we suppress operators of zero or negative integer dimension
which, by the analysis of \distler , correspond to BRST exact states.
Then we may write

\eqn\opexy{
\lim_{z\to w}{X(z)Y(w)} = \sum _i O_{j(i),0}(w)
,}

\noindent i.e. $X$ fused with $Y$ gives back other dimension zero $m$$=$$0$
physical state operators.

The fusion rules for $SU(1,1)/U(1)$ coset operators are greatly simplified
due to the absence of nontrivial $SU(1,1)$ null states in generic coset
modules. Usually in a Kac-Moody conformal field theory zeroes of the
$3$-point function arise from three distinct sources: vanishings of the
Wigner (Clebsch-Gordan)
coefficients associated with tensor products of the
underlying Kac-Moody primaries, vanishings due to null states in the
modules, and vanishings due to the conformal and global $SL(2,R)$
Ward identities\kz .
One writes:

\eqn\fusionrel{
O_{j_1,m_1} \circ O_{j_2,m_2} = \sum _j N(j,j_1,j_2)C^{j_1 j_2 j}
_{m_1 m_2 m_1+m_2} O_{j,m_1+m_2}
,}

\noindent where the $C$'s are the
Wigner coefficients. Now in the case at hand
we have not solved for the complete $3$-point functions (in part because
we do not know the proper definition of the $2$-point functions!) but
since we can compute Wigner coefficients we can determine most of
the fusion rules for the physical states of the string theory.

Thus, to determine which dimension zero operators appear on the
right-hand side of \opexy , we consider the classical $SU(1,1)$
tensor product

\eqn\tensorp{
\Phi _{j,0} = \sum ^\infty _{n=-1} C_n\, \Phi ^{lw}_{-5/4,n+3/4} \otimes
\Phi ^{hw}_{-5/4,-n-3/4}
,}

\noindent where we have abbreviated the Wigner coefficients by
$C_n$. Since $X$ and $Y$ are each descended from three distinct
Kac-Moody primaries, there are several other relevant tensor products,
but consideration of these does not alter the conclusion reached below.

Although the Wigner coefficients for tensor products of various
unitary representations of $SU(1,1)$ can be found in the mathematical
literature\puk ,\bieden , we have here a tensor product of lowest weight and
highest weight {\it nonunitary} representations. Thus it is safest to
proceed from first principles. A general discussion can be found in
appendix B.

The requirement that the right-hand side of \tensorp\ be an eigenstate
of the Casimir \casimir\ produces the following recursion relation:

\eqn\firstrec{
(n+\half )(n+2)C_{n+1}
+(n-\half )(n+1)C_{n-1}
-(j(j+1)+2n^2+3n+\half )C_n =0 .
}

\noindent This can be converted into a hypergeometric differential equation
by introducing the function

\eqn\dummyfn{
f(t) = \sum ^\infty _{n=0} C_{n-1}\, t^n .
}

\noindent Then \firstrec\ is solved by

\eqn\firstsol{
f(t) = (1-t)^j\, F(j-1/2, j+1; -1/2; t) .
}
The coefficients $C_n$ are obtained from \firstsol\ by expanding
the hypergeometric function in the standard hypergeometric series,
which converges at $t$$=$$1$ for $j< -1/2$. To determine for what
values of $j$ these Wigner coefficients are nonvanishing, we first
assume that the original tensor product states can be normalized (i.e.
that $X$ and $Y$ really exist in the WZW model). Then the Wigner coefficients
can self-consistently be taken as nonvanishing for those $j$ values
$\le -1/2$ such that

\eqn\wigser{
\sum ^\infty _{n=-1} |C_n|^2
,}

\noindent is a convergent series. This constraint has the unique solution

\eqn\jsol{
j=-3/2
\; .}

\noindent Actually the tensor product
\tensorp\ also contains the identity, which
is missed in the above argument because the identity is not normalizable
expressed in a normalized tensor product basis (i.e. $C_n = (-1)^n$).
This exception has been noted in the literature\bieden . So our final
result is the fusion rule

\eqn\xyrule{
X \circ Y \sim I + O^{cont}_{-3/2,0} .}

It is interesting to compare this result with the ground ring geometry
of the Liouville theory\ground . For the Liouville theory compactified
to the $SU(2)$ radius, this is a three dimensional cone coming from the
relation

\eqn\cone{
a_1a_2 - a_3a_4 =0
\; .}

\noindent Here we seem to have an analogous relation

\eqn\ourcone{
a_1a_2 + a_3a_4 = 1
\; .}

\noindent up to undetermined numerical coefficients. This is similar to
what is expected for the Liouville theory with nonzero cosmological
constant.

By considering arbitrary tensor products of $X$'s and $Y$'s one sees
that the fusion algebra generates all of the
physical dimension zero operators in the cohomology. For example, the
fusion of $X$ with itself gives

\eqn\xonx{
X \circ X \sim O^{lw}_{-3/2,3/2} + O^{double}_{1/2,3/2}
,}

\noindent where these grade two operators are defined modulo
normalizations by the expressions

\eqn\xsq{
\eqalign{
O^{lw}_{-3/2,3/2} = &\Bigl[ {3\over 10}\jt _{-2} -{1\over 5}
\jt _{-1}\jt _{-1} +{3\over 2}\j{+}_{-1}\j{-}_{-1}
\cr
&-{27\over 10}\j{+}_0\j{-}_{-2} +{39\over 5}\j{+}_{0}\jt _{-1}\j{-}_{-1}
+{99\over 8}\j{+}_0\j{+}_0\j{-}_{-1}\j{-}_{-1} \Bigr] \Phi ^{lw}_{-3/2,3/2}
,\cr
}
}

\noindent and

\eqn\dub{
O^{double}_{1/2,3/2}=\left[ 8\jt _{-1}\j{+}_{-1}
-21\j{+}_{-2}-15\j{+}_{-1}\j{+}_{-1}\j{-}_0 \right]
\Phi ^{double}_{1/2,1/2} .}
This result follows from the tensor product relations

\eqn\xxontod{
\Phi ^{lw}_{-5/4,-1/4} \otimes \Phi ^{lw}_{-5/4,-1/4}
= \Phi ^{double}_{1/2,-1/2}
,}

\noindent (with an unusual normalization) and

\eqn\xxalso{
\eqalign{
\Phi ^{lw}_{-5/4,-1/4}\otimes \Phi ^{lw}_{-5/4,7/4}
 - i\sqrt{2}\Phi ^{lw}_{-5/4,3/4}&\otimes\Phi ^{lw}_{-5/4,3/4}
- \Phi ^{lw}_{-5/4,7/4}\otimes\Phi ^{lw}_{-5/4,-1/4}
\cr
&=\Phi ^{lw}_{-3/2,3/2} ,\cr
}
}

\noindent (note that for this nonunitary tensor product the
Wigner coefficients cannot all be made real).

Not surprisingly, the $SL(2,R)$ ground ring is more complicated than
it's counterpart in $SU(2)$ Liouville. As we shall see, this is also
true of the symmetry algebra of physical currents.

\newsec{$W_\infty$ Algebra of the Black Hole $W$ hair}
\subsec{The charge algebra of the Liouville theory}

For the $SU(2)$ Liouville theory Klebanov and Polyakov\kleb\ and
Witten\ground\ have computed the algebra of the physical conserved
chiral charge operators

\eqn\chargedef{
Q^{\sss \pm}_{s,n} = \oint dz\, W^{\sss\pm}_{s,n}(z) .}

\noindent The algebra is remarkably simple:

\eqn\littlew{
\left[ Q^{\sss +}_{s,n} , Q^{\sss +}_{s^\prime ,n^\prime} \right]
= (n\sp  -n^\prime s)Q^{\sss +}_{s+s^\prime -1, n+n^\prime}
,}

\noindent and

\eqn\morew{
\eqalign{
\left[ Q^{\sss +}_{s,n} , Q^{\sss -}_{-s^\prime -1,n^\prime} \right]
&= -(n\sp -n^\prime s)Q^{\sss -}_{-s-\sp , n+n^\prime}
,\cr
\left[ Q^{\sss -}_{-s-1,n} , Q^{\sss -}_{-\sp -1, n^\prime} \right]
&= 0 .\cr
}
}

\noindent If we truncate the algebra to charges with $s$ integer, then
\littlew\ is precisely the wedge subalgebra of $w_{1+\infty}$,
the contraction of $W_\infty (\lambda )$ for $\lambda$$=$$1/2$
\pope ,\shen .
When a cosmological constant is added to the Liouville theory,
this symmetry algebra is modified to a much more complicated
uncontracted $W_\infty $ structure, which does not appear to
correspond to any of the previously catalogued algebras\barbon ,\dot .

We may take these facts as indications of what to expect for the
holomorphic part of the algebra of conserved charges in the $SL(2,R)$
black hole background. Since these charges represent
stringy ``hair'' of the black hole, what we are seeking may be termed
the algebra of $W$ hair of the two dimensional black hole\ellis .

In \littlew\ the $SU(2)$ $J^3_0$ eigenvalue $n$ is simply conserved
and plays the role of ``conformal'' mode number of the generators
of $w_{1+\infty}$. The same will be true for the $SL(2,R)$ charges

\eqn\ourcharge{
Q_{j,m} = \oint dz\, W_{j,m}(z)
,}

\noindent and thus we must obtain a wedge algebra from these charges
as well. Furthermore, upon
shifting the $SU(2)$ spin $s$ by $1$ in \littlew\ it becomes
equivalent to the ``spin'' of the corresponding $w_{1+\infty}$
generator; thus for example $Q^{\sss +}_{{\sss\pm}1}$, $Q^{\sss +}_{0}$
are the ``spin two''  Virasoro generators $L_{{\sss\pm}1}$, $L_0$ in
the sense of $W_\infty$. One might expect
that a similar relationship between $j$ and $s$ would hold for
the $SL(2,R)$ charge algebra, though this is not obvious.
It will turn out that for $Q_{j,m}$'s corresponding to continuous
series representations there is a simple relation between $j$ and
the $W_\infty$ ``spin'' $s$:

\eqn\spinrel{
s = -2j,
}

\noindent but $Q_{j,m}$'s corresponding to discrete series
representations do not have definite $s$.

Before delving into the details of the $W$ hair algebra, we want
to point out that \littlew\ may actually be a contraction of
a more general family of algebras than $W_{\infty}(\lambda )$.
To see this, consider the {\it super} $W_{\infty}(\lambda )$ algebra
of Bergshoeff et al\szeg . For $\lambda$$=$$0$ or $1/2$, it takes
the form\szeg :

\eqn\superalg{
\eqalign{
\left[ V^s_n , V^\sp _m \right] &= \sum ^\infty _{l=0} q^{2l}
g^{s\sp}_{2l}(n,m) V^{s+\sp -2l-2}_{n+m} + {\rm c.t.},
\cr
\left[ \vt ^s_n, \vt ^\sp _m \right] &= \sum ^\infty _{l=0} q^{2l}
\gt ^{s\sp}_{2l}(n,m) \vt ^{s+\sp -2l-2}_{n+m} + {\rm c.t.},
\cr
\left\{ \GP{s}_{n} , \GM{\sp}_{m} \right\} &=
\sum ^\infty _{l=0} q^l \left( b^{s\sp}_l(n,m) V^{s+\sp -l-1}_{n+m}
+ \bt ^{s,\sp}_l(n,m) \vt ^{s+\sp -l-1}_{n+m} \right)
+ {\rm c.t.},
\cr
\left[ V^s_n , \GPM{\sp}_m \right] &=
\sum ^\infty _{l=0} q^{l-1} (\mp 1)^{l+1} a^{s\sp}_l(n,m)
\GPM{s+\sp -l-1}_{n+m} ,
\cr
\left[ \vt ^s_n , \GPM{\sp}_m \right] &=
\sum ^\infty _{l=0} q^{l-1} (\mp 1)^{l+1} \at ^{s\sp}_l(n,m)
\GPM{s+\sp -l-1}_{n+m} ,
\cr
}
}

\noindent where c.t. stands for central terms. This algebra can be
contracted by defining\szeg

\eqn\newgen{
L^s_n = V^s_n + \vt ^s_n \; ,\quad \lt ^s_n = q( V^s_n - \vt ^s_n ),
}

\noindent then taking the limit $q$$\to$$0$. The contracted algebra in
this case is the $N$$=$$2$ super $w_\infty$ algebra\popeshen :

\eqn\littlesup{
\eqalign{
\left[ L^s_n , L^\sp _m \right] &= [n(\sp -1) - m(s-1)] L^{s+\sp -2}_{n+m},
\cr
\left[ L^s_n , \GPM{\sp}_m \right] &= [n(\sp -1) - m(s-1)]
\GPM{s+\sp -2}_{n+m},
\cr
\left[ L^s_n , \lt ^{\sp}_m \right] &=
[n(\sp -1) - m(s-1)] \lt ^{s+\sp -2}_{n+m} ,
\cr
\left[ \lt ^s_n , \GPM{\sp}_m \right] &=
\pm \GPM{s+\sp -1}_{n+m} ,
\cr
\left\{ \GP{s}_n , \GM{\sp}_m \right\} &=
-2L^{s+\sp -1} -2[n(\sp -1) -m(s-1)]\lt ^{s+\sp -2}_{n+m} ,
\cr
}
}

\noindent ($\lt$ is usually written as a current: $\lt ^s_n$$=$
$J^{s+1}_n$).

We speculate (but have not proven) that there exists a {\it bosonic}
version of the super $W_{\infty}(\lambda )$ algebra in which all generators
are bosonic and anticommutators are replaced by commutators. Further,
we imagine that the structure constants are unaltered except for those
changes necessary to recover an {\it automorphism} symmetry under
$\lambda$$\to$$\half$$-$$\lambda$, rather than the {\it anti-automorphism}
symmetry that determines much of the structure of super $W_{\infty}(\lambda )$
\szeg . Similar bosonic versions of superalgebras have been discussed
in the literature\romans ,\ber ,\ritten . Such an algebra, of course, has
only a superficial connection with genuine supersymmetry.

The $\lambda$$=$$1/2$ version of this bosonic superalgebra is obtained
from \superalg\ by replacing anticommutators with commutators and
letting

\eqn\newstruc{
\gt\to -\gt\; ,\quad \at\to -\at .
}
This algebra can be contracted by defining

\eqn\newnewgen{
L^s_n = q( V^s_n + \vt ^s_n )\; ,\quad \lt ^s_n = V^s_n -\vt ^s_n .
}

\noindent then taking the limit $q$$\to$$0$. The contracted algebra
has the same form as \littlesup , but with anticommutators replaced
by commutators, and with $L^s_n$$\leftrightarrow$$\lt ^s_n$. If we
now define

\eqn\qdefin{
Q^s_n = \half (\GP{s}_n + \GM{s}_n)
,}

\noindent we find the following contracted subalgebra:

\eqn\finalalg{
\eqalign{
\left[ \lt ^s_n , \lt ^\sp _m \right] &= [n(\sp -1) - m(s-1)]
\lt ^{s+\sp -2}_{n+m},
\cr
\left[ \lt ^s_n , Q^\sp _m \right] &= [n(\sp -1) - m(s-1)]
Q^{s+\sp -2}_{n+m},
\cr
\left[ Q^s_n , Q^{\sp}_m \right] &= [n(\sp -1) - m(s-1)]
\lt ^{s+\sp -2}_{n+m} .
\cr
}
}

\noindent Comparing \finalalg\
with \littlew , we see that \littlew , including the half-integer
spin generators, is precisely the wedge of \finalalg .
Furthermore, the $Q^{\sss -}$ generators of \morew\ are
precisely analogous to the {\it negative} spin $L^s_n$'s in this
contracted algebra.

\subsec{The $W$ hair algebra}

If we now examine the $SL(2,R)$ $W$ hair algebra, we will find an
uncontracted wedge of something similar to this bosonic superalgebra.
To determine commutators of the $SL(2,R)$ charges \ourcharge , we
need the $1/(z-w)$ terms in the operator products of the
corresponding $W_{j,m}(z)$'s (we ignore null states in this discussion).
Only the antisymmetric parts of these terms contribute to the commutators
of charges, however the symmetric parts are also needed in constructing
operators which are exactly marginal (see the following
section). Thus we are interested in the full ``lone star'' algebra\popeb\
not just the $W$ algebra.

We look first at operator products of continuous series $W_{j,m}$'s
with continuous series $W_{j,m}$'s fusing to other continuous series
$W_{j,m}$'s. These operator product coefficients are determined by
the Wigner coefficients for the appropriate tensor products of Kac-Moody
primary states at the base. Thus it suffices to examine tensor
products like

\eqn\genten{
\pc _{j,m_1+m_2} = \sum ^{\infty}_{n=-\infty} C_n
\pc _{j_1, m_1+n} \otimes \pc _{j_2,m_2-n} .
}

\noindent The sum is unrestricted since the tensor product is of
two continuous series reps. Furthermore since we are producing another
continuous series rep from the tensor product, there are no boundary
conditions to impose on solutions to the Wigner recursion relations.
In principle this means that there can be two independent solutions,
but in fact it suffices to exhibit the one given by

\eqn\bestsol{
C_n = \left[ {\Gamma (n-j_1+m_1) \Gamma (n+j_1+m_1+1) \over
\Gamma (n-j_2-m_2) \Gamma (n+j_2-m_2+1)} \right]^{1/2} D_n ,
}

\noindent where $D_{n-1}$ is the coefficient of $t^n$ in the
expansion of

\eqn\wholething{
f(t) = t^{(j_1-m_1+2)} (1-t)^{(j+m_1+m_2)}
F(j+j_1+j_2+2,j+j_1-j_2+1;2j_1+2;t) .}

\noindent One can easily see that the resulting Wigner coefficients
define a convergent series if and only if $j$ takes one of the values

\eqn\goodjs{
j = j_1+j_2+1,\; j_1+j_2+2,\; j_1+j_2+3,\,\ldots .
}

\noindent But in fact we have no solution at all, since for these
values $j$ and $m_1+m_2$ are either both integers or both half integers,
and thus do not parametrize a continuous series representation.
So there are no continuous series $W_{j,m}$'s at
all appearing in the product of two continuous series $W_{j,m}$'s.

There is one exception to this result, involving the operator
$\pc _{-1/2,0}$. Recall that this is the only non-square integrable
discrete state, and also corresponds to the only continuous series state
in a unitary representation. If we consider the operator product of
$\pc _{-1/2,0}$ with itself, the Wigner coefficients of the
classical tensor product are given by the expansion of

\eqn\hyperb{
f(t) = t^{3/2}(1-t)^j F(j+1,j+1;1;t) ,
}

\noindent which for $j$$=$$-1/2$ gives

\eqn\hyperc{
f(t) = t^{3/2}(1-t)^{-1/2}F(\half ,\half ,1,t) .
}

\noindent Comparing with \ccexp , we see that although \hyperc\
gives a divergent Wigner series, this appears to simply reflect
the fact that $\pc _{-1/2,0}$ is not square integrable, while
the original tensor product basis is, by definition, normalized.

Now let us determine which discrete series $W_{j,m}$'s appear
in the product of two continuous series $W_{j,m}$'s. The discrete
series $W_{j,m}$'s are always at sufficiently high grade that one
of the relevant tensor products will involve the lowest/highest weight
Kac-Moody primary at the base of the module. We write this state
as a tensor product of continuous series primaries:

\eqn\dtens{
\pmi _{-m_1-m_2,m_1+m_2} = \sum ^{\infty}_{n=-\infty}
C_n \pc _{j_1,m_1+n} \otimes \pc _{j_2,m_2-n} .
}

\noindent The Wigner coefficients are then trivially determined
by the condition that $J^{{\sss +}total}_0$ annihilate the state.
The unique solution is

\eqn\uniqsol{
C_n = (-1)^n \left[ {\Gamma (n+j_2-m_2+1) \Gamma (n-j_2-m_2) \over
\Gamma (n+j_1+m_1+1) \Gamma (n-j_1+m_1)} \right]^{1/2} .
}

\noindent The asymptotic behavior of these coefficients is

\eqn\asbeh{
\lim _{n\to\infty} C_n \sim n^{-m_1-m_2} = n^j ,
}

\noindent which thus give a convergent series for all discrete series
reps
with $j$$<$$-1/2$. Thus we reproduce the result of Puk\' anszky\puk\
that {\it all} unitary discrete
series reps appear in the product of {\it any two}
(Hermitian but not necessarily unitary) continuous series reps.
Of course, since all of our continuous series $W_{j,m}$'s have
$m$ values which are integer or half integer, we will only produce
the discrete series $W_{j,m}$'s which have $j$ integer or half integer.

Before interpreting these results in the language of $W$ algebras,
we treat one more case. Let us determine which continuous series
$W_{j,m}$'s appear in the product of a lowest weight $W_{j,m}$ with
a continuous $W_{j,m}$ (the case of highest weight $\times$ continuous
is precisely analogous). Thus consider

\eqn\lowcon{
\pc _{j,-j_1+m_2} = \sum ^{\infty}_{n=0} C_n \pmi _{j_1, -j_1+n}
\otimes \pc _{j_2,m_2-n} .
}

\noindent As before we solve for the Wigner coefficients:

\eqn\morewig{
C_n = \left[ {\Gamma (n-2j_1) \Gamma (n+1) \over
\Gamma (n-j_2-m_2) \Gamma (n+j_2-m_2+1)} \right]^{1/2} D_n ,
}

\noindent where the $D_n$ are obtained by expanding a dummy function
$f(t)$. Normally one would have to impose a boundary condition on $f(t)$
due to the fact that the $C_n$ series in \lowcon\ terminates at the lower
end (i.e. $n$$=$$0$); this boundary condition would be
$f(t)$$\to$$t$ as $t$$\to$$0$. However in the actual recursion relation
that defines the $C_n$, the $C_n$ for $n$$\ge$$0$ are decoupled from
the $C_n$ with $n$$<$$0$. Thus a consistent solution results from taking
any $f(t)$ (there are two independent solutions) to define the $C_n$ with
$n$$\ge$$0$, and taking the trivial solution $C_n$$=$$0$ for $n$$<$$0$.
We may thus find a solution from

\eqn\anotherf{
f(t) = t^{(2j_1+2)} (1-t)^{(j-j_1+m_2)}
F(j+j_1+j_2+2,j+j_1-j_2+1;2j_1+2;t) .
}

It is now easy to see that the resulting $C_n$'s define a convergent
series provided $j$ takes one of the values

\eqn\nicejval{
j = j_1+j_2+1,\; j_1+j_2+2,\; j_1+j_2+3,\,\ldots .
}

We can now recast these results in the language of $W_\infty$.
We use the correspondence \spinrel\ to write the continuous
series $W_{j,m}$'s as $W_\infty$ generators $L^s_n$:

\eqn\mapit{
\eqalign{
L^2_{{\sss\pm}1} &= \pc _{-1,{\sss\pm}3/2} \cr
L^3_{{\sss\pm}2} &= \pc _{-3/2,{\sss\pm}3} \cr
L^3_0   &=    \wc _{-3/2,0} \cr
L^4_{{\sss\pm}3} &= \pc _{-2,{\sss\pm}9/2} \cr
L^4_{{\sss\pm}1} &= \wc _{-2,{\sss\pm}3/2} \cr
&\ldots\cr
}
}

\noindent Note that these are all of the integer spin $W_\infty$
generators in the wedge which have $s$$+$$n$ odd. The exception is
$L^1_0$, which is a special case; if we defined this to be
simply $\pc _{-1/2,0}$, then it would not behave at all like a
$U(1)$ current, since

\eqn\ponp{
\pc _{-1/2,0} \circ \pc _{-1/2,0} \sim \pc _{-1/2,0}
+ \wp _{-1,0} +\wm _{-1,0} +\, \ldots .
}

\noindent So instead we {\it define} $L^1_0$ to be the exactly marginal
operator which can be constructed as an infinite series beginning with
$\pc _{-1/2,0}$:

\eqn\uonecur{
\eqalign{
L^1_0 &= \pc _{-1/2,0} + \wp _{-1,0} - \wm _{-1,0} +\, \ldots ,
\cr
L^1_0 &\circ L^1_0 = 0 .\cr
}
}
Note in the above expression (and many that follow) we suppress
overall numerical coefficients which could anyway be absorbed
into normalizations; however an
important relative minus sign was made explicit in \uonecur .

Furthermore, the results derived above imply that we can obtain
all of the remaining
integer spin $W_\infty$ generators in the wedge as {\it infinite
sums} of discrete series $W_{j,m}$'s. For example, we can define
$L^2_0$ from the operator product

\eqn\gimelo{
\eqalign{
L^2_1 \circ L^2_{-1} &= \pc _{-1,3/2} \circ \pc _{-1,-3/2}
\cr
&\sim \wp _{-1,0} +\wm _{-1,0} +\wp _{-2,0} +\wm _{-2,0} +\, \ldots
\cr
&\equiv L^2_0 .
}
}

\noindent Given this we may then define $L^4_0$ from the operator
product

\eqn\gimefou{
\eqalign{
L^4_1 \circ L^4_{-1} &= \wc _{-2,3/2} \circ \wc _{-2,-3/2}
\cr
&\sim \wp _{-1,0} +\wm _{-1,0} +\wp _{-2,0} +\wm _{-2,0} +\, \ldots
\cr
&\equiv L^4_0 + L^2_0 .
}
}

\noindent The symmetries of the Wigner coefficients under
Weyl reflection are such that the series of operators defining
$L^4_0$ will begin with $\wp _{-2,0}$ and $\wm _{-2,0}$.

Continuing with our examples:

\eqn\gimelthon{
\eqalign{
L^3_2 \circ L^2_{-1} &= \pc _{-3/2,3} \circ \pc _{-1,-3/2}
\cr
&\sim \wp _{-3/2,3/2} + \wm _{-3/2,3/2}
+\wp _{-5/2,3/2} +\wm _{-5/2,3/2} +\, \ldots
\cr
&\equiv L^3_1
}
}

We will assume that the infinite series appearing in this construction
produce normalizable and orthogonal states (it would be better to prove
this). In that case, given our results and the simple rules for
products of unitary discrete series representations\bieden ,
we have obtained an uncontracted integer spin $W_\infty$ wedge algebra:

\eqn\ourform{
\left[ L^s_n , L^\sp _m \right] =
\sum ^{\infty}_{l=0} g^{s\sp}_{2l}(n,m)L^{s+\sp -2l-2}_{n+m}
,}
where the structure constants $g^{s\sp}_{2l}(n,m)$ are computable
--with rapidly increasing difficulty-- from conformal field theory
techniques. It should be emphasized that the difficulty in computing
structure constants derives solely from the fact that the current
algebraic construction of the higher grade discrete states
themselves becomes rapidly very complicated.

We have not yet accounted for all of the independent combinations
of discrete series $W_{j,m}$'s with $j$ integer or half integer.
These can be interpreted once we consider the $W_{j,m}$'s with
$j$ quarter integer. These operators are all discrete series, and
correspond to half-integer spins $s$$=$$3/2$, $5/2$, $7/2$, ... in
our mapping to $W_\infty$. The spin $3/2$ generators have the
following products:

\eqn\spinalgops{
\eqalign{
\pc _{-3/4,3/4} \circ \pc _{-3/4,3/4} &= \wm _{-3/2,3/2} ,
\cr
\pc _{-3/4,-3/4} \circ \pc _{-3/4,-3/4} &= \wp _{-3/2,-3/2} ,
\cr
\pc _{-3/4,3/4} \circ \pc _{-3/4,-3/4} &= 0 .
}
}

\noindent The full product algebra generated from these spin
$3/2$ operators appears to account precisely for the ``extra''
half-integer and integer spin generators. Furthermore, comparing
\spinalgops\ with \littlesup , we see that this extra structure gives
our $W$ hair algebra the form of a bosonic superalgebra as described
in general terms above. Since the generators of the $W$ hair algebra
act on the ground ring, which itself is generated by $j$$=$$1/4$
operators $X$ and $Y$, we speculate that our algebra may be
related to the ``symplecton'' of Biedenharn and Louck\louck .

In closing this section, we consider the question of why the embedding
of the $W_\infty$ discrete state algebra is so much more complicated
for the $SL(2,R)$ black hole than for flat space Liouville theory.
The answer is that $SU(2)$ Liouville
is a very special case. The $w_{1+\infty}$
algebra is simply related to an $SU(2)$ enveloping algebra, and
thus fits very neatly into the underlying $SU(2)$ Kac-Moody current
algebra. For the black hole the $W_\infty$ algebra of the
discrete states has no apparent relation to the underlying $SL(2,R)$
Kac-Moody current algebra, even though this current algebra has its
own $W_\infty$ structure\kir . On the other hand, we can now, in
hindsight, understand better why the discrete states obtained
by Distler and Nelson had to be such a motley assortment of
$SL(2,R)$ states. Nonunitary continuous series reps appear
because unitary continuous reps would fuse back onto the continuum
of unitary continuous reps\puk . Unitary discrete series reps appear
because these fuse onto an infinite series of other unitary discrete
reps with increasing $|j|$, just as we get from fusing two continuous
series reps. Nonunitary discrete series reps appear in the ground ring
states because they fuse to a series of discrete reps with
decreasing $|j|$.


\newsec{Deformations of the Black Hole Background}

The dimension $(1,1)$ (marginal)
operators of the $SL(2,R)/U(1)$ conformal field theory
\eqn\moduli{
O(z,\bar z) = W_{jm} \bar W_{jm^{\prime}}\; ,
}
are the infinitesimal moduli for the black hole background
and can generate deformations of the theory that preserve the central charge.
If such an operator retains its conformal dimension in the {\it deformed}
theory it generates a one-parameter family of conformally invariant
backgrounds of fixed central charge and is called {\it exactly} marginal.

The maximal set of mutually commuting currents of a Kac-Moody algebra are
independent exactly marginal operators, since the
operator product of any two such currents
does not contain any simple pole pieces which would give logarithmic
contributions to the two-point function and thus shift the conformal dimension
of the current \kadanoff\chsch . For example, in the $c=1$ model at the
$SU(2)$ radius there is only one exactly marginal
operator, $J_3 \bar J_3$, which
changes the radius of the target space \radius . In the Liouville theory, the
algebra of the discrete states is $w_{1+\infty}$ so that the set of
independent exactly marginal operators
is the subset $\wwp _{s,0}\bar\wwp _{s,0}$,
$s$$=$$0,1, \ldots $.  In the case at hand,
the discrete series $W_{j,m}$ do not behave like
``currents'' in that their
operator products with themselves produces an infinite number
of other discrete series $W_{j,m}$'s. As discussed in the previous
section, we expect rather that the physical moduli are constructed
from an abelian subset of the $W_\infty$ generators. In particular,

\eqn\modguys{
\eqalign{
&L^1_0\bar L^1_0 \; ,\cr
&L^2_0\bar L^2_0 \; ,\cr
}
}
are exactly marginal deformations of the $SL(2,R)$ black hole
background.

It is possible to compute explicitly,
in the semi-classical ($k$$\to$$\infty$) limit\wit ,
the back reaction on the black
hole background from such an operator.
We construct the appropriate composite field in the ungauged $SL(2,R)$ WZW
model using the results of Section 3, and gauge the axial diagonal $U(1)$.
Such a gauge invariant operator can be added to the action of the gauged
$ SL(2,R)$ WZW model,
and one can compute the infinitesimal deformation of the
background to lowest order in $1/k$.
These deformations may then be compared with approximate solutions of
the beta function equations for the black hole in non-trivial tachyon
backgrounds \dealwis ,\mandal ,\rabi .

To discuss target space physics we use the Euler angle parametrization of
the WZW fields \translate . In unitary gauge,
$\theta_L$$=$$-\theta_R$$=$$\theta $,
this gives
\eqn\gaugefix{
g_{\pm\pm} = \co r ~, ~~~ g_{\pm\mp}
= \si r ~e^{{\sss\mp}i\theta}\, ,
}
and the action of the gauged $SL(2,R)$ WZW model is
\eqn\witac{
L= {{k }\over {2 \pi}} \int d^2z ~ \partial_z r \partial_{\bar z} r
+ \si ^2 r
\left ( \partial_z \theta \partial_{\bar z } \theta +
2 A_z \partial_{\bar z } \theta -
2 A_{\bar z} \partial_z \theta \right) - 4 \co ^2 r A_z A_{\bar z}
}
The Kac-Moody primaries \hyper-\jzero\ are readily translated into unitary
gauge and the currents \currents\ can be written as
\eqn\currents{\eqalign{
\j{\pm}(z) &= \pm \kappa ~(\; - \partial_z r
\pm {{i}\over{2}} ~\si 2r ~ \partial_z \theta ~)~
e^{{\sss\mp}i
\theta} \cr
\jb{\pm}(\bar z) &= \mp \kappa ~(\; - \partial_{\bar z} r
\mp {{i}\over{2}} \si 2r ~\partial_{\bar z} \theta ~)~
e^{{\sss\pm} i \theta}  \cr
}
}
The discrete states come in pairs related by a {\it duality} symmetry
which flips the sign of the right-handed (holomorphic) $U(1)$ current.
Although the string effective action is expected to be duality invariant to
all orders, the deformations of the background need not respect duality
symmetry.

Using the results of Section 3, it is easy to show that the composite fields
\eqn\states{
\eqalign{
\psi^{\sss\pm\pm} &= : \left( \jb{\pm}\right)^N \left(
\j{\pm}\right) ^N \left ( \gpp \right )^{j+m-N} : \cr
\psi^{\sss\pm\mp} &= : \left( \jb{\pm}\right) ^N \left( \j{\mp}
\right) ^N \left ( \gpm \right )^{j+m-N} : \cr}
}
are discrete states lying on the
boundary of a highest/lowest weight module
at grade $N$ with coset dimension $(1,1)$, and which obey the physical
state conditions \mshell . Their $(j,m)$ values correspond to $s$$=$$1$,
$r$$=$$1$, $2$,
$\ldots $ in \hwphys-\lwphys .
For grade three and upwards, there are additional discrete states living
in the bulk of the module which can be constructed with more difficulty.
They correspond to $s>1$ in \hwphys-\lwphys .

The WZW composite fields are covariantized according to the
prescription:
\eqn\covg{
\eqalign{
\gb (z,\bar z) &\rightarrow e^{{{i}\over{2}}
\sigma_2 \int^z A_z d z }~ \gb (z,\bar z)
{}~ e^{{{i}\over{2}}\sigma_2 \int^{\bar z } A_{\bar z } d \bar z }
\cr
\partial_z \gb  &\rightarrow  \partial_z \gb + {{i}\over{2}}
A_z ( \sigma_2~ \gb + \gb ~ \sigma_2 )  .\cr }
}
One can show that functionals of $\gpp$, $\gpm$ with
non-vanishing $J^3(\bar J^3)$ eigenvalue are dressed by Wilson lines.
Also, ordinary derivatives of functionals with $m$$+$$\bar m \not= 0$ are
converted into covariant derivatives. For example, the derivatives of the
spinor weights transform as
\eqn\wilson{
\eqalign{
\partial_z \gpp &\rightarrow
e^{\pm \left({{i}\over{2}} \int^z A_z d z + {{i}\over{2}}\int^{\bar z}
A_{\bar z} d \bar z \right) } \left (\partial_z \gpp \pm 2~ i~ A_z
\gpp \right )\cr
\partial_z \gpm &\rightarrow
e^{\pm \left({{i}\over{2}} \int^z A_z d z - {{i}\over{2}}\int^{\bar z}
A_{\bar z} d \bar z \right) }
\left( \partial_z \gpm \pm 2~i~A_z \gpm \right) .
\cr
}
}
The currents $\j{\pm}$, $\jb{\pm}$ of the ungauged $SL(2,R)$ model become
parafermions of the coset model
\eqn\parafermi{
\eqalign{
\Psi^{\sss\pm} &=e^{{\sss\pm} i \int^z A_z d z }
\left( \gpm \partial_z
\gpp \pm 2~ i~ A_z \gpp  - \gpp \partial_z \gpm \right)
\cr
\bar\Psi^{\sss\pm} &=e^{{\sss\pm} i \int^{\bar z} A_{\bar z} d \bar z }
\left(
\gmp \partial_{\bar z} \gpp \pm 2~ i~ A_{\bar z} \gpp  -
\gpp \partial_{\bar z} \gmp \right)
\cr
}}
These expressions are reminiscent of the ``classical parafermions'' in
\bardacki .

There are four independent grade one moduli of the black hole.
Each of them has
$m$$=$$\bar m$$=$$0$. In this case the vector potential
is particularly easy to integrate out of the action since it only appears
in the covariant derivatives. Their explicit form is
\eqn\gradea{
\eqalign{
\psi^{\sss ++} + \psi^{\sss --} &= -
2~\se ^2 r \partial_z r \partial_{\bar z} r  - 2~\si ^2 r \left(
\partial_z \theta
\partial_{\bar z} \theta
+ 2~ A_z \partial_{\bar z} \theta - 2 ~ A_{\bar z} \partial_z \theta
- 4 ~ A_z A_{\bar z} \right)
,\cr
\psi^{\sss ++} - \psi^{\sss --} &= 2 i~ \ta r \left ( \partial_z \theta
\partial_{\bar z} r + \partial_z r \partial_{\bar z} \theta \right )
  - 4~ i ~\ta r \left (
A_z \partial_{\bar z } r + A_{\bar z } \partial_z r \right )
,\cr
}
}
and their partners under the duality transformation:
$\psi^{\sss +-} \pm \psi^{\sss -+}$.

We should also consider the deformation produced by \ccexp :

\eqn\cosmo{
\Phi^{c-c}_{-1/2,0,0}(r) =  {{1}\over{cosh ~r }}F \left (
{{1}\over{2}}, {{1}\over{2}}; 1; tanh^2 r \right )
={{1}\over{cosh ~ r }} K(tanh^2 r)
}
where $K$ is an elliptic function. Recall that in the Liouville
theory the analogous operator is the zero mode of the tachyon, and
is exactly marginal.
In our case the deformation \cosmo\ also corresponds to turning on
a nonzero tachyon background, however \cosmo\
is {\it not} exactly marginal.
Rather, as we have discussed, the exactly marginal deformation
which turns on a nonzero tachyon background is

\eqn\realthing{
L^1_0\bar L^1_0 =
\Phi ^{c-c}_{-1/2,0,0} +
i (\psi^{\sss ++} - \psi^{\sss --}) + \ldots .
}
which includes back-reaction on the space-time metric
as well as higher order corrections involving the
``massive modes'' of the string.
Since the Lagrangian is quadratic and non-derivative in $A$ we can
integrate out the vector potential giving
\eqn\otachy{\eqalign{
L = \partial_z r \partial_{\bar z } r &( 1 + 4 \alpha^2 \ta ^2 r ~ \se ^2 r )
+ 2 \alpha ~ \ta r ~ \se ^2 r (\partial_z r \partial_{\bar z } \theta
+ \partial_z \theta \partial_{\bar z } r ) \cr
 &+ \ta ^2 r \partial_z \theta
\partial_{\bar z } \theta  + \alpha \Phi ^{c-c}_{-1/2,0,0}(r) ,\cr}
}
where $\alpha$ an arbitrary parameter.
We diagonalize the metric via the coordinate transformation
\eqn\rc{
\theta \rightarrow \tilde \theta (r,\theta) ~, ~~~~ {{\partial \tilde \theta}
\over {\partial \theta }} = 1 ~, ~ ~ {{\partial \tilde \theta}\over
{\partial r }} = 2 \alpha ~\cs r ~ \se r ,
}
which gives
\eqn\newmetric{
G_{rr} = 1 - 4 \alpha^2 \se ^2 r
{}~, ~~~ G_{\tilde \theta \tilde \theta} = \ta ^2 r .
}
To compare with the beta function results we change variables,
$ \ta ^2 r = 1- \mu e^{-2\rho} $,
so that the metric takes the form
\eqn\tachymet{
ds^2 = {{1-4 \alpha^2 \mu e^{-2 \rho} }\over{1-\mu e^{-2\rho}}}
( d \rho)^2 + (1-\mu e^{- 2 \rho} ) (d \tilde \theta)^2
\, ,}
and the one-loop contribution, from the integration over $A$, is the linear
dilaton background
\eqn\ldil{
\Phi(r)=\lo \co r \rightarrow 2 \rho - \lo \mu\;  .}

Now we recall
that the beta function equations are
only solved in the weak
field approximation for the tachyon, which corresponds to large $r$.
Then we note that, as
$r$$\to$$\infty $, the elliptic function in \cosmo\
tends to a logarithm\abrom
\eqn\whatiz{
K(\ta ^2r ) \rightarrow \lo \co r + 2 \sqrt 2
\, ,}
so that the
static tachyon background is
\eqn\tachy{
T(\rho) \sim \alpha (\rho - {{1}\over{2}} \lo \mu ) \sqrt \mu e^{-2 \rho}
\, .}
These expressions are in
complete agreement with eq.(15) of \dealwis ,
where the back reaction on the black hole metric was computed
from the beta function equations. The
parameter $\mu$ plays the role of the mass of the black hole. The Hawking
temperature can be calculated as in \dealwis .

Finally, consider perturbing the action of the gauged WZW model by the
duality invariant operator
\eqn\ohawk{
\eqalign{
\alpha O(z,\bar z) &= \alpha \left (\psi^{\sss ++}
+ \psi^{\sss --} +  \psi^{\sss +-} +
\psi^{\sss -+} \right )\cr
 & =-2\alpha \left(\se ^2 r+\cs ^2 r \right)\partial_z r \partial_{\bar z} r
+ 2 \alpha \partial_z \theta \partial_{\bar z } \theta
- 4\alpha \left ( A_z \partial_{\bar z} \theta - A_{\bar z } \partial_z \theta
\right ) + 8\alpha  A_z A_{\bar z} .\cr}
}
This is the simplest part of the exactly marginal deformation
$L^2_0\bar L^2_0$.
Integrating out $A$ as before yields
\eqn\mess{
L = \partial_z r \partial_{\bar z} r \left[ 1- 2 \alpha (\cs ^2 r + \se ^2 r )
\right] + \partial_z \theta \partial_{\bar z} \theta \left[ \si ^2 r
+ 2 \alpha - {{ (\si ^2 r +
2 \alpha)^2}\over {\co ^2 r + 2 \alpha }} \right] .}
The one-loop contribution to the action from the measure in the integration
over $A$ gives the target space dilaton term with
\eqn\dilaton{
\Phi = \lo [ \co ^2 r + 2 \alpha]
}
To compare with the original black hole metric we now make a
coordinate transformation that converts the dilaton back to its original
form
\eqn\coord{
\co ^2 r + 2 \alpha  \rightarrow \co ^2 r
}
and to $O(\alpha)$, we obtain the target space metric
\eqn\metric{
ds^2 = {{k}\over{2}} \left (~(dr)^2 + \ta ^2 r ~(d \theta)^2
{}~\right)
}
Thus we find that,
to lowest order in the coupling, $\alpha$, the perturbation
simply rescales the original action by an overall constant.

\newsec{Conclusion}

It must certainly be possible to put the $SL(2,R)/U(1)$ coset
string theory on a firmer formal footing. We have avoided some
technical issues in this paper, not because they are intractable,
but rather because they can be more confidently addressed
once the underlying physical content has been made manifest.
We believe that we have made substantial progress in this direction.
It would be well to prove unitarity and modular invariance
for this theory, and to resolve the remaining ambiguity in the
definition and evaluation of physical correlators.

We would also like to get more information about the $W$ hair
algebra and its precise relation to other $W_\infty$ structures.
Although we have made some speculative remarks in this regard,
a much better job can and should be done.

As we have observed, the physical $W$ hair structure has great
difficulty embedding itself in the underlying abstract $SL(2,R)$
Kac-Moody algebra. This may be a hint that coset conformal field
theories are not very well suited for describing stringy black
holes, and that we will be better off in the long run turning to
alternate methods. It may well be that these structures are more
accessible in other formulations such as string field theory.

\vskip .3in
\noindent {\bf Acknowledgments:}\ We would like to thank J. Cohn
and J. Distler for helpful discussions.

\appendix{A}{Explicit Construction of $SL(2,R)$ Primaries}

In the $SL(2,R)$ WZW theory the Kac-Moody primaries are composites
of $\gp$, $\gm$, $\fm$, and $\fp$. Their form is completely determined
up to normalizations by the requirement that they form irreducible
representations of classical $SU(1,1)_L$$\times$$SU(1,1)_R$. The
$g$'s transform like $j$$=$$\bar\jmath$$=1/2$ spinors, with $(m,\bar m)$
components $(1/2,1/2)$, $(-1/2,-1/2)$, $(-1/2,1/2)$ and $(1/2, -1/2)$,
respectively. We write a general ansatz for a primary in the form

\eqn\genfop{
\Phi _{a,b,c,d}(z,\bar z) = \sum ^{\infty}_{n=-\infty}
R_n (\gp )^{a+n}(\gm )^{b+n}(\fm )^{c-n}(\fp )^{d-n}
,}
where the $R_n$ are coefficients and the parameters $a$, $b$, $c$, $d$
satisfy

\eqn\kolp{
a+b+c+d=2j\; .
}
Note that $j$$=$$\bar\jmath$ is automatic in this construction.
Following Vilenkin\vilenkin , it is useful to define a rescaled
function

\eqn\resca{
\Phi _{a,b,c,d} (z,\bar z) = |\fm |^{2j} \; \pt _{a,b,c,d}
\left( {\gp\over |\fm |},{\gm\over |\fm |},e^{i\ph}\right) ,
}
where exp$\,i\ph$ denotes $\fm /|\fm |$.

Now $\pt$ must be an eigenstate of both $\js _L$ and $\js _R$ with
eigenvalue $-j(j+1)$. This gives a recursion relation for the
coefficients $R_n$:

\eqn\whatr{
(a+n+1)(b+n+1)R_{n+1} + (c-n+1)(d-n+1)R_{n-1}
+[(a+n)(b+n)+(c-n)(d-n)]R_n = 0 .
}
This is solved by

\eqn\rrrs{
R_n = {(-1)^n \over \Gamma (a+n+1) \Gamma (b+n+1)
\Gamma (c-n+1) \Gamma (d-n+1)} .}
We thus obtain

\eqn\hereyougo{
\Phi _{a,b,c,d} =
\sum ^{\infty}_{n=-\infty} {
(\gp )^a (\gm )^b (\fm )^c (\fp )^d
(-\fp\fm /\gp\gm )^n \over
\Gamma (a-n+1) \Gamma (b-n+1) \Gamma (c+n+1) \Gamma (d+n+1) }
,}
and from the action of the raising and lowering operators we
can identify the parameters $a$, $b$, $c$, $d$ as

\eqn\whattheyis{
a+d = j+m\, ,\; b+c = j-m\, ,\; a+c = j+\bar m\, ,\;
b+d = j-\bar m .
}

It is clear by inspection of \hereyougo\ that the parameter $d$
is redundant and can be set to zero. We then obtain, using a
simple choice of normalization, the expression \hyper .

\appendix{B}{$SU(1,1)$ Tensor Products}

$SU(1,1)$ tensor products have the form

\eqn\thisfo{
\Phi _{j,m_1+m_2} = \sum ^{\infty}_{n=-\infty} C^{j_1j_2j}_{m_1m_2m_1+m_2}(n)
\Phi _{j_1,m_1+n} \otimes \Phi _{j_2-n}
,}
where we assume that the individual tensor product states are normalized.
The requirement that $\Phi _{j,m_1+m_2}$ be an eigenstate of
the total $\js $ gives a second order recursion relation for the
Wigner coefficients:

\eqn\grecurs{
\eqalign{
0=
&~C_{n+1} \sqrt{(j_1-m_1-n)(j_1+m_1+n+1)(j_2+m_2-n)(j_2-m_2+n+1)}
\cr
&+C_{n-1} \sqrt{(j_1+m_1+n)(j_1-m_1-n+1)(j_2-m_2+n)(j_2+m_2-n+1)}
\cr
&+C_n \bigg[ -j(j+1)+(m_1+m_2)(m_1+m_2-1)
\cr
&\qquad +(j_1-m_1-n+1)(j_1+m_1+n)
+(j_2+m_2-n)(j_2-m_2+n+1) \bigg] .\cr
}
}
In the special case that $\Phi _{j,m_1+m_2}$ is supposed to be either
highest or lowest weight, then we may use a simpler first order
recursion relation. For example, requiring that $\Phi _{j,m_1+m_2}$ be
annihilated by $\j{-}_1 + \j{-}_2$ gives

\eqn\firo{
C_{n+1} \sqrt{(j_1-m_1-n)(j_1+m_1+n+1)} + C_n
\sqrt{(j_2-m_2+n+1)(j_2+m_2-n)} = 0
,}
which is solved trivially.

In the more general case of \grecurs , it is useful to define new
coefficients by

\eqn\newcos{
C_n = \left[ {\Gamma (n-j_1+m_1) \Gamma (n+j_1+m_1+1)
\over \Gamma (n-j_2-m_2) \Gamma (n+j_2-m_2+1) } \right]^{1/2} D_n .}
The new coefficients $D_n$ satisfy

\eqn\drec{
\eqalign{
0=
&~(n-j_1+m_1)(n+j_1+m_1+1)D_{n+1} +(n-j_2-m_2-1)(n+j_2-m_2)D_{n-1}
\cr
&-\bigg[ j(j+1) - (m_1+m_2)(m_1+m_2-1)
\cr
&\qquad +(n-j_1+m_1-1)(n+j_1+m_1)
+(n-j_2-m_2)(n+j_2-m_2+1) \bigg] D_n .\cr
}
}
To solve this, we introduce a dummy function $f(t)$:

\eqn\dumfun{
f(t) = \sum _{n} D_{n-1} t^n .}
One can then show that $f(t)$ satisfies Riemann's differential
equation in the form\morse

\eqn\riem{
f(t) = P\pmatrix{0&1&\infty \cr
                 \lambda &\mu & \nu \cr
                 \lambda ^{\prime} & \mu ^{\prime} & \nu ^{\prime} \cr}
,}
where

\eqn\pardes{
\eqalign{
\lambda &= -j_1-m_1+1\; ,\quad \lambda ^{\prime} = j_1-m_1+2 \cr
\mu &= j+m_1+m_2\; ,\quad \mu ^{\prime} = -j-1+m_1+m_2 \cr
\nu &= j_2-m_2\; ,\quad \nu ^{\prime} = 1-\lambda -\lambda ^{\prime}
-\mu -\mu ^{\prime} -\nu .\cr
}
}
One solution of this equation is

\eqn\onesolis{
f(t) = t^{\lambda} (1-t)^{\mu} F(a,b;c;t)\, ,
}
where

\eqn\abcdare{
\eqalign{
a&=\lambda +\mu +\nu \cr
b&=1-\lambda ^{\prime} -\mu ^{\prime} -\nu \cr
c&=1-\lambda ^{\prime} +\lambda .
\cr
}
}

In general, there is also a second independent solution. This is
given by interchanging $\lambda$ and $\lambda ^{\prime}$ in
\onesolis\ and \abcdare . Note that

\eqn\noth{
a+b-c = 2j-1 .
}
If in addition $c$ is not zero or a negative integer, then this
implies that, for both solutions, the
corresponding hypergeometric series
is convergent at $t$$=$$1$ for all $j$$<$$-1/2$. This means
that we can evaluate the hypergeometric functions by their
series expansions

\eqn\hserex{
F(a,b;c;1) = {\Gamma (c)\over \Gamma (a)\Gamma (b)}
\sum ^{\infty}_{n=0} {\Gamma (n+a)\Gamma (n+b) \over
\Gamma (n+c) \Gamma (n+1) } .}
We may thus determine the coefficients $D_n$ and $C_n$.

To determine which $j$ values actually occur in the tensor product,
we must first impose any boundary conditions on the solution which
will arise if any of $\Phi _{j}$, $\Phi _{j_1}$, or $\Phi _{j_2}$
is a discrete or double-sided representation. We must also
require

\eqn\notinfg{
\sum _n |C_n|^2 < \infty\;
,}
so that $\Phi _{j,m_1+m_2}$ is normalizable.

In addition to the examples worked out in the text, we note below some
useful results for tensor products of discrete series reps.
\item{--} The tensor product of two unitary highest weight reps
contains only other highest weight reps.
The only possible $j$ values
$j_1+j_2$, $j_1+j_2-1$, $j_1+j_2-2$, $\ldots$.
\item{--} The same is true for the product of two unitary lowest
weight reps.
\item{--} In the tensor product of a highest weight and lowest weight
representation with the same $j$, the only discrete rep which appears
is the identity\bieden .

\listrefs
\listfigs
\vglue .5in
\vbox{\tabskip=0pt \offinterlineskip
\def\tablerule{\noalign{\hrule}}
\halign to5.5in{\bigstrut#& \vrule#\tabskip=1em plus2em&
  \hfil#& \vrule#& \hfil#\hfil& \vrule#&
  \hfil#& \vrule#\tabskip=0pt\cr\tablerule
&&\omit\hidewidth Liouville\hidewidth&&
 \omit\hidewidth $SL(2,R)$\hidewidth&&
 \omit\hidewidth Grade\hidewidth\nl
&&$\wwp _{0,0}$&&$\pc _{-1/2,0}$&&0\nl
&&$\wwp _{1/2,{\sss\pm}1/2}$&&$\Phi ^{hw/lw}_{-3/4,{\sss\mp}3/4}$&&$0$\nl
&&$\wwp _{1,0}$&&$\wp _{-1,0}$, $\wm _{-1,0}$&&$1$\nl
&&$\wwp _{1,{\sss\pm}1}$&&$\pc _{-1,{\sss\pm}3/2}$&&$0$\nl
&&$\wwp _{3/2,{\sss\pm}3/2}$&&$\Phi ^{hw/lw}_{-5/4,{\sss\mp}9/4}$&&$0$\nl
&&$\wwp _{3/2,{\sss\pm}1/2}$&&$W^{hw/lw}_{-5/4,{\sss\pm}3/4}$&&$2$\nl
&&$\wwp _{2,{\sss\pm}2}$&&$\pc _{-3/2,{\sss\pm}3}$&&$0$\nl
&&$\wwp _{2,{\sss\pm}1}$&&$\wp _{-3/2,{\sss\pm}3/2}$, $\wm
_{-3/2,{\sss\pm}3/2}$
&&$3$\nl
&&$\wwp _{2,0}$&&$\wc _{-3/2,0}$&&$4$\nl
&&$\wwp _{5/2,{\sss\pm}5/2}$&&$\Phi ^{hw/lw}_{-7/4,{\sss\mp}15/4}$&&$0$\nl
&&$\wwp _{5/2,{\sss\pm}3/2}$&&$W^{hw/lw}_{-7/4,{\sss\pm}9/4}$&&$4$\nl
&&$\wwp _{5/2,{\sss\pm}1/2}$&&$W^{hw/lw}_{-7/4,{\sss\mp}3/4}$&&$6$\nl
&&$\wwp _{3,{\sss\pm}3}$&&$\pc _{-2,{\sss\pm}9/2}$&&$0$\nl
&&$\wwp _{3,{\sss\pm}2}$&&$\wp _{-2,{\sss\pm}3}$, $\wm
_{-2,{\sss\pm}3}$&&$5$\nl
&&$\wwp _{3,{\sss\pm}1}$&&$\wc _{-2,{\sss\pm}3/2}$&&$8$\nl
&&$\wwp _{3,0}$&&$\wp _{-2,0}$, $\wm _{-2,0}$&&$9$\nl
}}
\vfil\centerline{TABLE 1}\eject
\vglue .5in
\vbox{\tabskip=0pt \offinterlineskip
\def\tablerule{\noalign{\hrule}}
\halign to5.5in{\bigstrut#& \vrule#\tabskip=1em plus2em&
  \hfil#& \vrule#& \hfil#& \vrule#\tabskip=0pt\cr\tablerule
&&\omit\hidewidth Ground Ring States\hidewidth&&
 \omit\hidewidth Physical States\hidewidth\nl
&&$\od _{-1,0}$ (identity)
&&$W^{hw/lw}_{-1,0}$, $\pc _{-1,{\sss\pm}3/2}$\nl
&&$O^{hw/lw}_{-5/4,{\sss\mp}3/4}$
&&$\Phi ^{hw/lw}_{-5/4,{\sss\mp}9/4}$,
$W^{hw/lw}_{-5/4,{\sss\pm}3/4}$\nl
&&$O^{hw/lw}_{-3/2,{\sss\mp}3/2}$, $\od _{1/2,{\sss\pm}3/2}$,
$\oc _{-3/2,0}$
&&$\pc _{-3/2,{\sss\pm}3}$,
$\wp _{-3/2,{\sss\pm}3/2}$, $\wm _{-3/2,{\sss\pm}3/2}$,
$\wc _{-3/2,0}$\nl
&&$O^{hw/lw}_{-7/4,{\sss\mp}9/4}$, $O^{hw/lw}_{-7/4,{\sss\pm}3/4}$
&&$\Phi ^{hw/lw}_{-7/4,{\sss\mp}15/4}$,
$W^{hw/lw}_{-7/4,{\sss\pm}9/4}$, $W^{hw/lw}_{-7/4,{\sss\mp}3/4}$\nl
}}
\vfil\centerline{TABLE 2}\eject
%
\bye